\documentstyle[procl, twoside]{article}

\input{psfig.sty}

\def\Journal#1#2#3#4{(#1) {#2} {\bf #3}, #4}

\def\AAp{\em Astron. Astrophys.}
\def\AApL{\em Astron. Astrophys., Lett.}
\def\AAps{\em Astron. Astrophys., Suppl. Ser.}
\def\AJ{\em Astron.~J.}
\def\ApJ{\em Astrophys.~J.}
\def\ApJL{\em Astrophys.~J., Lett.}
\def\ApJSS{\em Astrophys.~J., Suppl. Ser.}

\def\MNRAS{\em Mon. Not. R.~Astron. Soc.}

\def\PASP{\em Publ. Astron. Soc. Pac.}

\def\etal{{et al. \rm}}
\def\FE{[Fe II] $\lambda$1.644 $\mu$m}
\def\FES{[Fe II] $\lambda$1.644 $\mu$m }

\def\fe{[Fe II] }

\def\ls{$L$(\FE) }
\def\l{$L$(\FE)}
\begin{document}

\markboth{T. Morel, R. Doyon, \& N. St-Louis}{Near-IR \fe Line Imaging Survey of SNRs in M33}

\thispagestyle{plain}

\title{A Near-IR \fe Line Imaging Survey of Supernova Remnants in M33}

\author{Thierry Morel,$^1$ Ren\'e Doyon,$^2$ \& Nicole St-Louis$^2$}

\address{$^1$ Astrophysics Group, Imperial College of Science, Technology and Medicine, Blackett Laboratory, Prince Consort Road, London, SW7 2BZ, UK; t.morel@ic.ac.uk.\\
$^2$ D\'epartement de Physique, Universit\'e de Montr\'eal, 
C. P. 6128, Succ. Centre-Ville, Montr\'eal, Qu\'ebec, Canada, H3C 3J7; and
Observatoire du Mont M\'egantic; doyon, stlouis@astro.umontreal.ca.}

\maketitle

\abstract{We report on the first near-IR \fe line imaging survey of extragalactic SNRs. Observations of a sample of 42 objects drawn from an optically-selected catalogue of SNRs in M33 provide evidence for a wide range in the \linebreak \FES luminosities. 
This can be understood as being primarily due to variations in the chemical abundances and density of the local ISM, although shock conditions may also play a significant role. 
We briefly discuss how these results may be used to better calibrate the supernova rate of star-forming galaxies.}

\section{Why this survey?}

The use of the strong near-IR \fe lines as a mean of estimating the supernova rate of starburst galaxies has become increasingly popular over the last decade (e.g., Calzetti 1997). However, one serious limitation of this technique lies in the fact that one has to adopt typical values for the \fe luminosity and [Fe II]-emitting lifetime of a {\em single} supernova remnant (SNR), whereas a wide span of values are observed. For instance, the \fe luminosities of SNRs in the starburst galaxy M82 (Greenhouse \etal 1997) are two orders of magnitude higher than in ``quiescent'' galaxies, such as our Galaxy or the LMC (Oliva \etal 1989). One order of magnitude higher values are also observed in NGC 253 (Forbes \etal 1993). Therefore, before applying this technique to star-forming galaxies that presumably exhibit widely different properties (e.g., metallicity), one must first understand what parameters control the \fe properties of individual SNRs. In an effort to address this issue, we report here on preliminary results of a \FES line imaging survey of an heterogeneous sample of SNRs in M33. 

\section{Observations}
Our observations were obtained during two observing runs at the Canada-France-Hawaii telescope ({\em CFHT}) in 1997 and 1998. Two different infrared
cameras were used: {\em MONICA} in 1997 and {\em REDEYE} in 1998. A total of 42 objects drawn from the optically-selected atlas of Gordon \etal (1998) has been observed. Particular attention has been paid to observe SNRs with widely different properties. The standard reduction procedure for infrared imaging has been applied (see, e.g., Hodapp \etal 1992).

\section{Relationship between the \fe emission and  other SNR properties}
The broad range of \FES luminosities is made clear by the fact that --- while several SNRs present an outstanding level of \fe emission (see Fig.1) --- many of them remain undetected in our survey.

\begin{figure}[htb]
\centerline{\psfig{figure=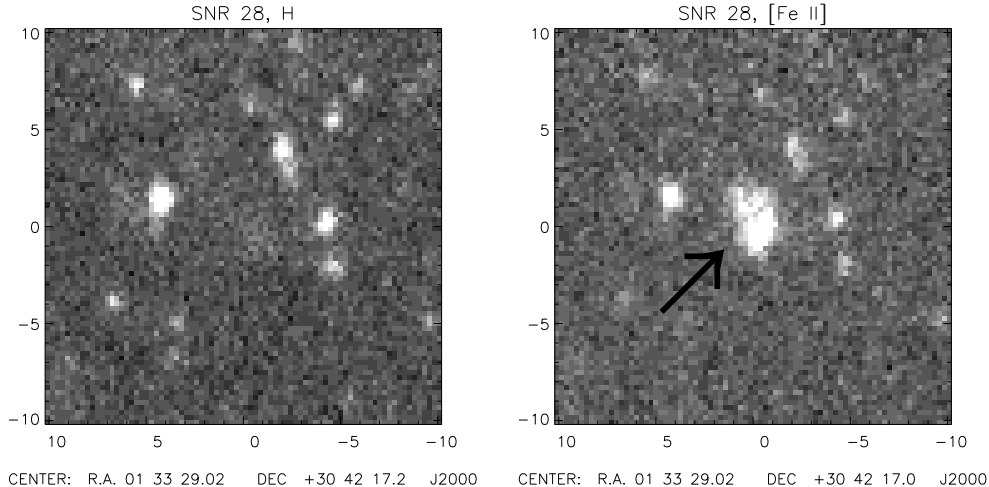,width=14truecm,%
       bbllx=35pt,bblly=470pt,bburx=564pt,bbury=740pt}}
\caption{Comparison of the $H$ ({\it left panel}) \rm and \FES ({\it right panel}) \rm images of SNR 28 (identification from Gordon \etal 1998). The SNR is indicated by an arrow. The field of view is 20$''$ $\times$ 20$''$. North is up, and east is to the left. The integration times are 300 s and 1 500 s for {\it H} and [Fe II], respectively.}
\end{figure}

The \FES luminosities (or upper limits) are presented in Table 1. It can be seen that a (nearly) two orders of magnitude range is observed in these values. Lumsden \& Puxley (1995) reported near-IR spectroscopic observations of 9 of our program targets. There is a good agreement between the two estimates.\footnote{Their [Fe II] $\lambda$1.257 $\mu$m fluxes have been converted into \FES fluxes assuming a ratio: [Fe II] $\lambda$1.257 $\mu$m/\FES = 1.36 (Nussbaumer \& Storey 1988).}

\rm \begin{table}[h]
\caption{\FES luminosities of SNRs in M33.$^a$}
\begin{center}
\hspace*{-1cm} 
\begin{tabular}{cccccccc}
\cline{1-2}\cline{4-5}\cline{7-8}
SNR Id.$^b$  & $L$ (\FE)   && SNR Id.$^b$ & $L$ (\FE)   && SNR Id.$^b$  & $L$ (\FE)   \\
                 &(L$_{\odot}$)&&                 &(L$_{\odot}$)&&                  &(L$_{\odot}$)\\
\cline{1-2}\cline{4-5}\cline{7-8}
    &                 &&    &                 &&    &             \\
9   & 59.9 $\pm$ 37.2 && 25 & 70.6 $\pm$ 28.6 && 50 & $<$ 432     \\
10  & $<$ 203         && 26 & $<$ 374         && 54 & $<$ 92.2    \\
12  & $<$ 189         && 28 & 475 $\pm$ 105   && 55 & 280 $\pm$ 98\\
13  & $<$ 61.0        && 29 & $<$ 94.1        && 57 & $<$ 225     \\	
14  & $<$ 300         && 30 & $<$ 163         && 58 & $<$ 228     \\	
15  & $<$ 694         && 31 & 671 $\pm$ 208   && 60 & $<$ 19.5    \\  	
16  & $<$ 1039        && 32 & $<$ 836         && 66 & $<$ 616     \\  
17  & $<$ 651         && 34 & $<$ 207         && 73 & $<$ 337     \\
18  & $<$ 53.8        && 35 & 695 $\pm$ 217   && 74 & $<$ 1040    \\
20  & $<$ 17.8        && 38 & $<$ 85.8        && 80 & $<$ 306     \\
21  & $<$ 174         && 41 & $<$ 2349        && 82 & $<$ 230     \\
22  & $<$ 335         && 43 & $<$ 192         && 86 & $<$ 244	\\
23  & $<$ 1142        && 44 & $<$ 142         && 87 & $<$ 103	\\
24  & $<$ 72.3        && 45 & $<$ 116         && 94 & 189 $\pm$ 51\\\cline{1-2}\cline{4-5}\cline{7-8}\\
\end{tabular}
\end{center}
$^a$ Assuming a distance to M33 of 840 kpc (Freedman, Wilson, \& Madore 1991). The upper limits are quoted for a statistical significance of 3$\sigma$. The error bars for the detected objects are 1$\sigma$ uncertainties.\\ 
$^b$ Identifications from Gordon \etal (1998).\\
\end{table}

In order to pin down the physical processes that produce the \fe emission, we sought for correlations between the \FES luminosities and various quantities intrinsic to the SNRs, such as their age or optical-line properties, for instance. We made use of the generalized Kendall's tau correlation technique which allows us to investigate statistical correlations between datasets including upper limits (Isobe, Feigelson, \& Nelson 1986). 
Table 2 summarizes the results of this statistical analysis for all quantities under consideration, along with the source of these data.  

\rm \begin{table}[h]
\caption{Results of the statistical analysis. $N$: Number of points included in the calculations. ${\cal P} (X, Y)$: Probability for a chance correlation between the variables $X$ and $Y$. Note that the objects at large galactocentric distances, $GCD$, show a tendency for {\em lower} \ls values.}
\begin{center}
\begin{tabular}{cccc}
\hline
$X$  &  $Y$                                 & $N$     & ${\cal P} (X, Y)$ \\\hline
     &                                      &         &                   \\
\ls  & $GCD$$^a$                            & 42      & 8.5 \%\\
$''$ & $n_e$$^a$                            & 34      & 0.1 \%\\
$''$ & Dynamical age$^a$                    & 42      & 82 \%\\
$''$ & Metal abundance$^b$                  & 20      & 0.2 \%\\
$''$ & $V_{bulk}$$^c$                       & 8       & 8.3 \%\\
$''$ & $\cal{F}$ ([S II] $\lambda$6717)$^d$ & 34      & 0.1\%\\
$''$ & $S_6$$^e$                            & 35      & 1.5 \%\\
$''$ & $S_{20}$$^e$                         & 35      & 2.2 \%\\
$n_e$$^a$ & $S_{20}$$^e$                    & 34      & 5.6 \%            \\\hline\\
\end{tabular}
\end{center}
$^a$: From Gordon \etal (1998). The dynamical ages have been derived using their quoted optical diameters.\\
$^b$: From Blair \& Kirshner (1985) and Smith et al. (1993).\\
$^c$: From Blair, Chu, \& Kennicutt (1988).\\
$^d$: From Smith et al. (1993), Gordon \etal (1998), and Blair \& Kirshner (1985).\\
$^e$: From Duric et al. (1993) and Gordon et al. (1999).
\end{table}

In a preliminary effort to calculate the dynamical age of the SNRs, we have simply assumed that they all undergo Sedov-Taylor expansion, as suggested by Gordon \etal (1998). Without a detailed knowledge of the evolutionary status, initial blast energy of the SNR, and of the ambient gas density in which it evolves, age estimates are notoriously uncertain. Therefore, the complete lack of any relationship between the level of \fe emission and the dynamical age of the remnant (see Table 2) should be treated at this stage of the data analysis with caution. 

Our data support the idea that environmental rather than evolutionary effects are more \linebreak important in controlling the \fe properties. A tight correlation is found between the electronic densities derived from the [S II] $\lambda$$\lambda$6717, 6731 doublet and \l. This confirms that the \linebreak \fe emission arises in the dense postshock region.
The brightest \fe remnants are characterized by $n_e$([S II]) $\approx$ 10$^3$ cm$^{-3}$, which translates to $n_e$([Fe II]) $\approx$ 10$^{3.5}$ cm$^{-3}$ (e.g., Lumsden \& Puxley 1995). This requires preshock densities significantly higher than canonical values for the ISM ($n_e$ $\approx$ 0.1-10 cm$^{-3}$), approaching values typical of molecular material. Radio emission from SNRs is due to synchrotron emission, and is thus primarily a function of the number of available electrons and of the magnetic field strength. The tendency for the objects with high radio fluxes to be strong \fe emitters may thus be secondary in nature, and merely reflect the fact that dense environments both enhance the \fe and radio emission (provided that the magnetic properties are roughly similar within the sample). There is indeed some indication of a correlation between $n_e$ and $S_{20}$ (Table 2).   

A tight correlation is also observed between \ls and the generalized ``metal abundance'', $A$, as defined by Dopita \etal (1984). This most likely indicates that \fe emission is enhanced in regions characterized by high ISM chemical abundances. Because of the substantial metallicity gradient of M33 (Garnett \etal 1997), we should therefore observe a strong negative correlation between \ls and the SNR galactocentric distance. However, such evidence is lacking in our data (false alarm probability of 8.5 \%), probably owing to the large scatter in the GCD-metallicity relation (Smith et al. 1993). Although iron-bearing grains are easily destroyed by shocks (Draine \& Salpeter 1979), both theory and observations suggest that the efficiency of this process may vary from one SNR to another (Jones, Tielens, \& Hollenbach 1996; Oliva \etal 1999). The observed correlation between {\em A} and \ls  may also suggest that \fe emission is favoured by strong shocks that return a large fraction of the dust grains into the gas phase. In this case, we should expect a correlation between the \FES luminosities and the expansion velocity $V_{bulk}$ (to first order proportional to the shock speed). This correlation is also not clear in our data, possibly because our range of observed $V_{bulk}$ is not broad enough. Different electronic densities can also introduce noise in the correlation.

Although \FES is believed to probe somewhat cooler, denser material than the optical [S II] doublet (e.g., Oliva \etal 1989), the correlation between the \FES and [S II] fluxes suggests that the line-formation regions of these two species are not drastically different and that similar physical processes give rise to both emissions. All optical-line fluxes correlate extremely well with \l.

\section{Discussion}

In the light of the results presented above, we conclude that density and (possibly to a less extent) abundance effects are likely to play a pivotal role in controlling the \fe emission in SNRs. This may help to understand the differences that exist between the near-IR \fe properties of SNRs in galaxies with moderate star-formation activity (such as M33) and starburst galaxies. While the present study demonstrates that the brightest remnants have luminosities of the order of 700 L$_{\odot}$, SNRs in the starburst galaxy M82 have luminosities up to 1.6 $\times$ 10$^5$ L$_{\odot}$ (Greenhouse \etal 1991, 1997). Although it remains to be seen whether the two samples trace two similar populations of SNRs, we propose that this dichotomy may be explained by the dense, dusty environments prevailing in the nuclear regions of these galaxies. In particular, it is conceivable that processing (via grain destruction) of the large reservoir of dust grains greatly enhances the \fe emission with respect to more dust-free galaxies.


\section*{References}\noindent

\references

Blair, W. P., Chu, Y. -H., Kennicutt, R. C. (1988), in {\em Supernova Remnants and the Interstellar Medium}, eds. R. S. Roger \& T. L. Landecker, Proc IAU Colloq. 101, Cambridge Univ. Press, Cambridge, p.~193

Blair, W. P., \& Kirshner, R. P. \Journal{1985}{\ApJ}{289}{582}.

Calzetti, D. \Journal{1997}{\AJ}{113}{162}

Dopita, M. A., Binette, L., D'Odorico, S., \& Benvenuti, P. \Journal{1984}{\ApJ}{276}{653}

Draine, B. T., \& Salpeter, E. E. \Journal{1979}{\ApJ}{231}{438} 

Duric, N., Viallefond, F., Goss, W. M., \& van der Hulst, J. M. \Journal{1993}{\AAps}{99}{217}


Forbes, D. A., Ward, M. J., Rotatiuc, V., Blietz, M., Genzel, R., Drapatz, S., van der Werf, P. P, \& Krabbe, A. \Journal{1993}{\ApJL}{406}{L11}

Freedman, W. L., Wilson, C. D., \& Madore, B. F. \Journal{1991}{\ApJ}{372}{455}

Garnett, D. R., Shields, G. A., Skillman, E. D., Sagan, S. P., \&
Dufour, R. J. \Journal{1997}{\ApJ}{489}{63}

Gordon, S. M., Kirshner, R. P., Long, K. S., Blair, W. P., Duric, N., \& Smith, R. C. \Journal{1998}{\ApJSS}{117}{89}

Gordon, S. M., Duric, N., Kirshner, R. P., Goss, W. M., \& Viallefond, F. \Journal{1999}{\ApJSS}{120}{247}

Greenhouse, M. A., Woodward, C. E., Thronson Jr., H. A., Rudy,
R. J., Rossano, G. S., Erwin, P., \& Puetter, R. C. \Journal{1991}{\ApJ}{383}{164}

Greenhouse, M. A., et al. \Journal{1997}{\ApJ}{476}{105}

Hodapp, K.-W., Rayner, J., \& Irwin, E. \Journal{1992}{\PASP}{104}{441}

Isobe, T., Feigelson, E. D., \& Nelson, P. I. \Journal{1986}{\ApJ}{306}{490} 

Jones, A. P., Tielens, A. G. G. M., \& Hollenbach, D. J. \Journal{1996}{\ApJ}{469}{740}

Long, K. S., Charles, P. A., Blair, W. P., \& Gordon,
S. M. \Journal{1996}{\ApJ}{466}{750}


Lumsden, S. L., \& Puxley, P. J. \Journal{1995}{\MNRAS}{276}{723}

Nussbaumer, H., \& Storey, P. J. \Journal{1988}{\AAp}{193}{327}

Oliva, E., Moorwood, A. F. M., \& Danziger,
I. J. \Journal{1989}{\AAp}{214}{307}

Oliva, E., Lutz, D., Drapatz, S., \& Moorwood, A. F. M. \Journal{1999}{\AApL}{341}{L75} 

Smith, R. C., Kirshner, R. P., Blair, W. P., Long, K. S.,
\& Winkler, P. F. \Journal{1993}{\ApJ}{407}{564}

\end{document}